\documentstyle[iopfts]{ioplppt}
\begin{document}
\title{Circles in the Sky: Finding Topology with the Microwave Background
Radiation}

\author{Neil J. Cornish$^{\diamondsuit}$, David N. Spergel$^{\dag}$
and Glenn D. Starkman$^{\ddag}$}

\address{$\diamondsuit\,$Department of Applied Mathematics and Theoretical
Physics, University of Cambridge, Silver Street, Cambridge CB3 9EW, UK}
\address{\dag\ Department of Astrophysical Sciences, Princeton
University, Princeton, New Jersey 08544, USA}
\address{\ddag\ Department of Physics, Case Western Reserve
University, Cleveland, Ohio 44106-7079, USA}

\begin{abstract}
If the universe is finite and smaller than the distance
to the surface of last scatter, then the signature of the 
topology of the universe is writ large on the microwave background sky.
We show that the microwave background will
be identified at the intersections of the surface of last scattering
as seen by different ``copies'' of the observer.  Since the surface of
last scattering is a two-sphere, these intersections will be circles,
regardless of the background geometry or topology.
We therefore propose a statistic that is sensitive to all small, locally
homogeneous topologies. Here, small means that the distance to the
surface of last scatter is smaller than the ``topology
scale'' of the universe.  

\end{abstract}

\section{Introduction}
One of the goals of research in cosmology is
to answer basic questions about the universe:
``What is its structure?''
``Is it infinite or finite?" 
``Will it last forever?'' and ``How will it end?''.  
In the context of general relativity, 
these questions can be stated more formally as 
``What is the {\it geometry} and {\it topology} of the universe?''

If the universe is homogeneous and isotropic on large scales,
then its {\it geometry} is determined entirely by $\Omega$, 
the ratio of the current average energy density 
to the critical energy density. If $\Omega>1$, 
then the geometry of the universe is positively curved, 
like the surface of a sphere; 
the volume of the universe is finite;
and, for most equations of state,
the universe will ultimately recollapse in a Big Crunch.
If $\Omega=1$, 
then the geometry is flat, like a sheet of paper, 
and the universe will go on expanding forever,
albeit at a velocity that asymptotically approaches zero.
Finally, if $\Omega<1$, 
then the geometry is hyperbolic (negatively curved)
like the surface of a saddle,
and the universe will go on expanding forever,
at a velocity that does not asymptotically approach zero.

{\it Geometry} constrains, but does not dictate, {\it topology}.
If the geometry of the universe is flat, 
then it can either be infinite or compact.  
There are different compact universes
associated with each crystal group: 
for example, a three torus corresponds to a cubic symmetry.   
On the other hand, if the geometry of the universe
is positively curved ($\Omega > 1)$,
then the universe must be compact.
Finally, if the universe is hyperbolic, 
then again it can be either infinite or compact.  
There is a rich branch of mathematics
associated with the study of compact hyperbolic geometries~\cite{thurston}.

There are several physical and philosophical motivations for
considering compact universes.
Einstein and Wheeler advocate finite universes
on the basis of Mach's principle~\cite{Mach}.
Others argue that an infinite universe
is unaesthetic and wasteful~\cite{genetic}
because anything that can happen does happen, 
and an infinite number of times.
Quantum cosmologists have argued~\cite{nucleation,hypergenesis}
that small volume universe also have small action
and are therefore more likely to be created.
More intuitively,
it is difficult to produce a large universe,
so it happens less often.
Finally, a common feature of many quantum theories of gravity
is the compactification of some spacelike dimensions.
This suggests a dimensional democracy, in which
all dimensions (or at least all space-like dimensions) are compact.
Some dimensions remained at Planck or GUT scales,
while the three we observer grew to macroscopic proportions.

Most of the scant attention to non-trivial compact topologies in cosmology 
has focused on the simplest non-trivial topology of the flat geometry:
a cube with opposite sides identified,
{\it i.e.}~ a rectangular three-torus, $T^3$.
While the universe may be truly flat ($\Omega \equiv 1$, not
just $\vert \Omega - 1\vert \ll 1$), there is then no scale set by the
geometry, so the dimensions of the fundamental cell of the topology 
(the radii of the torus) are arbitrary.
It would be an unexpected and  unnecessary coincidence
if one of those scales was exactly of order the horizon size today.

If $\Omega<1$, however, then there is a natural
scale for the topology, namely the curvature scale.
Indeed, the compact topologies of $H^3$ (hyperbolic 3-d
space) are classified by their volume in units
of the curvature scale.  It has been shown~\cite{minvol}
that the volume of any compact hyperbolic 3-manifolds
is bounded below by $V_{{\rm min}}=0.166\, R_{curv}^3$,
and many explicit examples have been constructed with small volumes.
A collection of relatively simple topologies have been constructed
by identifying the faces of the four hyperbolic analogs of the
Platonic solids,
the hexahedron, icosahedron, and two dodecahedra~\cite{best}.
These examples typically have volumes in the range $(4-8)\, R_{curv}^3$,
but other simple examples have volumes of as small as
$0.94\, R_{curv}^3$\cite{weeks}. With the advent of computer aided
topology, such as the publicly available {\em SnapPea}
program~\cite{snappea}, there are now thousands of explicit examples known
with volumes less than $10  R_{curv}^3$.
It has also been show~\cite{thurston} that all three manifolds are built of
primitives which are homeomorphic (topologically equivalent) to one of
eight possible manifolds of constant geometry.  
Moreover, in a well-defined mathematical sense, 
{\em most} three manifolds are homeomorphic to manifolds of constant
negative curvature, {\it i.e.} to topologies of $H^3$.

Recently, we~\cite{cornish} proposed a new model for a compact
hyperbolic inflationary universe. This model was motivated by
observations that suggest $\Omega < 1$ (see the review by D. Spergel
in this volume\cite{david}). 
Previous attempts to construct hyperbolic inflationary
models\cite{openinf} assumed that the universe was infinite and
required various fine tunings to avoid producing enormous microwave
fluctuations on large angular scales. By assuming that the universe
was hyperbolic and compact, we were able to solve the large-scale
isotropy and homogeneity problems as long as the volume of the
universe was not much larger than $R_{curv}^3$, where
$R_{curv}=H_0^{-1}(1-\Omega_0)^{-1/2}$ is the curvature scale.
In this expression $H_0 = h 100$ km/s/Mpc is the Hubble constant and
$\Omega_0$ is the matter energy density today in units of the critical
density $\rho_{c}=3H_0^2/8\pi G$.

\section{Generic Features of Topology}
Whether the geometry be flat or hyperbolic,
there are certain characteristic signatures
of topology which can be looked for.
The surface of last scattering (SLS) is a two-sphere of radius
\begin{equation}
R_{sls}= R_{curv} \, {\rm arccosh}\left( {
2-\Omega_0 \over \Omega_0}\right) \, ,
\end{equation}
from which the cosmic microwave background radiation (CMBR)
photons were emitted.
In the limit $\Omega_0 \rightarrow 1$ the above expression reduces to
$R_{sls}(\Omega_0=1)= 2 c H_0^{-1} \simeq 6000 h^{-1}$ Mpc. 

In most cosmological models, microwave fluctuations on large angular
scales are due to variations in the gravitational potential
at the surface of last scatter. This is true in open inflationary
universes on angular scales in the range $\theta_{H} < \theta <
\theta_{curv}$ where $ \theta_H\simeq \sqrt{\Omega_0}\, 0.9^o $
is the angle subtended by the Hubble patch at last scatter and
$\theta_{curv} \simeq { (\Omega_0 / \sqrt{1-\Omega_0}}) \, 96^o$
is the angle subtended by the curvature scale on the surface of last
scatter. On scales smaller than $\theta_H$ the microwave fluctuations
are amplified by plasma oscillations, while on scales comparable to
and larger than $\theta_{curv}$ the fluctuations are enhanced by the
decay of the curvature perturbations along the line of sight\cite{ks}.
Across the range $\theta_{H} < \theta <\theta_{curv}$, experiments
such as COBE-DMR can be thought of as mapping the gravitational
potential along the inner surface of a two sphere whose radius
is $R_{sls}.$ If the physical dimension of the universe is less than
the diameter of the sphere of last scatter, $D_{sls}=2 R_{sls}$,
then the sphere of last scatter crosses back on itself and self-intersects.
The loci of self-intersections are circles. Thus fluctuations in the
CMBR would be correlated around pairs of circles with the same radii,
centred on different points on the sky. 

The self intersection of the SLS is easiest to visualise
from the perspective of the universal covering space, rather from the
confines of the topology's fundamental cell\footnote{A simple example
is the two dimensional torus $T^{2}=S^{1}\times S^1 = E^2/\Gamma$.
The universal cover is Euclidean flat space, $E^2$, and the fundamental
cell is either a right rectangle or a hexagon with opposite sides identified.
The fundamental cell tiles the covering space. The fundamental group
$\Gamma$ consists of discrete translations.}. Fixing our attention on a
constant time spatial hypersurface, there will be copies or clones of
ourselves dotted about the universe at positions dictated by the
fundamental group. Surrounding each clone will be a copy
of the sphere of last scatter. If any of our clones are situated a distance
less than $D_{sls}$ away from us, then our sphere of last scatter
will intersect the clone's. Of course, there is no physical
distinction between
us and our clones, so the intersections seen in the covering
space are in fact {\em self intersections} of the sphere of last
scatter. Taking some artistic licence, the picture in Fig.~1 shows
a topology where there are four clones within a radius of $D_{sls}$
of us. Each intersection produces a pair of correlated circles on the
sky -- one pair for each clone inside $D_{sls}$. In the example
shown, each circle pair has a different radius. Moreover, the two
circle pairs coming from the clones to the upper right of the
picture intersect each other.

As we shall discuss below, the existence of  these  correlated circles 
allows us to search for the existence of topology in general,
independent of the particular topology in question.  
It is important to emphasise here that {\em the 
signature is not constant temperature along each circle, 
but identical temperatures at identified points lying along pairs of
circles}.

\newpage
\
\begin{figure}[h]
\vspace{70mm}

\includegraphics{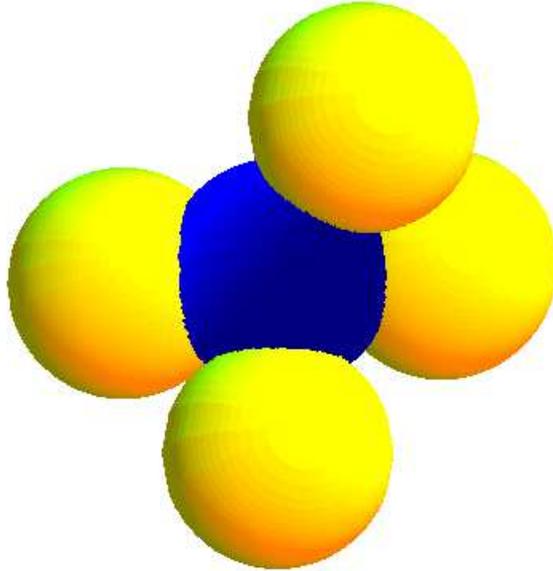}

\vspace{2mm}
\caption{The sphere of last scatter viewed in the universal cover.
The dark sphere marks the primary copy and the four
lighter spheres are intersecting clones.}
\end{figure}

Previous attempts to detect topology in a finite universe 
have used statistics 
that are only sensitive to $T^3$ topologies\cite{SSS,deOliveira}.
In contrast, our method is generic to all topologies that are locally
homogeneous and isotropic (this includes all small FRW models). Importantly,
the mapping from the surface of last scatter to the night sky 
is a conformal map. Since conformal maps preserve angles, 
the identified circles at the surface of last scatter 
would appear as equally rescaled, identified circles on the night sky.
The angular radius and angular separation of each identified circle pair
will depend on the geometry and topology of the universe, 
as will the number of pairs. If we are able to detect these circles,
then their position, number and size can
be used to determine the geometry and topology of the universe.

A second generic feature of topology is that 
it makes space globally anisotropic.
This can be understood quite simply in the case of a three-torus 
in which looking along one of the axes brings you back around in a
closed loop, but looking off-axis makes you wind round and round
the space like the red strip around a barber pole.
Thus, in a non-trivial topology, there are preferred directions.
What is more surprising is that all but $T^3$ also make space 
globally inhomogeneous. In most topologies, 
the identifications of faces are made with twists 
(much like how a Mobius strip is constructed from a length of ribbon).
Thus typical isometries involve a corkscrew type motion.
Since the topologies violate global isotropy, 
this mixing of translations and rotations causes a violation
of global homogeneity.  

Unlike other inhomogeneous cosmologies, such as Tolman-Bondi universes
which are locally inhomogeneous,  these global violations of
homogeneity and isotropy are not excluded. After all, 
we already know  that the universe is weakly inhomogeneous and
anisotropic on large scales -- there is observable structure. 
Similarly, in the topologically interesting cosmologies,
homogeneity and isotropy are violated only by the correlations between
the structure that we observe -- such as the fluctuations in the CMBR.

In principle, the locations of the identified circles
can also be used to determine the orientation and location
of the observer within the topology. More precisely, the topology
can equally well be described in terms of many different fundamental cells,
which share the same group structure, but represent
different choices of generators for the fundamental group.
The shape of the fundamental cell which an observer infers
will depend on the observer's location within the universe.

\subsection{Searching for Circles in the Sky}

Since the basic signature of topology is identified circles on
the night sky, we have developed a statistical tool to detect
these circles in all-sky maps of the CMBR. We begin by
selecting two points on the night sky, $\overrightarrow{x},$ and $
\overrightarrow{y}.$ We then draw circles of angular radius $\alpha$ around
each point and consider all possible relative phases, $\phi_* ,$ 
between the two circles. Defining $T_1(\phi)$ and $T_2(\phi+\phi_*)$ to be
the temperature fluctuations around the circles centred at
$\overrightarrow{x}$ and $\overrightarrow{y}$ respectively, we define
the circle comparison statistic
\begin{equation}\label{cstat}
S(\phi_*) = { <2 T_1(\pm \phi) T_2(\phi+\phi_*)> \over <T_1(\phi)^2
+T_2(\phi+\phi_*)^2>} \, ,
\end{equation}
where $<>=\int_0^{2\pi} d\phi$. The statistic ranges in the interval
$[-1,1]$. Circles that are perfectly matched have $S=1$, while an
ensemble of uncorrelated circles will have a mean value of $S=0$. By
searching for circles that are either anti-phased or phased, we are able to
detect both orientable and non-orientable topologies. Matched circles
in orientable topologies have clockwise-anticlockwise temperature
correlations while non-orientable topologies have a mixture of
clockwise-clockwise and clockwise-anticlockwise correlations.

In any practical application, the number of truly independent measurements
around each circle will be finite. An experiment with
angular resolution $\Delta\theta$ provides roughly
$N\simeq 2\pi \sin\alpha / \Delta\theta$ data points around each
circle of angular radius $\alpha$. Neglecting the
galaxy cut, the COBE-DMR instrument
yields $N\simeq 36 \sin\alpha$ pixels around each circle. In
comparison, MAP should yield $N\simeq 1800 \sin\alpha$ measurements
around each circle using its highest frequency channel.

For circles which are identified due to topology, and which have their
relative phase properly adjusted, $T_1$ should equal $T_2$ at each
point around the circles. However, noise causes there to be a
non-zero probability that $S\neq 1$. It is a reasonable
approximation to treat the noise in each data pixel as
an independent gaussian random variable with variance $\sigma_\eta^2$:
\begin{equation}
P(\eta) = {1\over \sigma_\eta \sqrt{2\pi}}e^{-\eta^2/2\sigma_\eta^2} .
\end{equation}
Assuming $T_1(\phi)=T(\phi)+\eta_1(\phi)$ and
$T_2(\phi)=T(\phi)+\eta_2(\phi)$, where $\eta_i(\phi)$ is the noise,
it is possible to derive an expression for the probability
distribution of $S$ for matched circle pairs. The result is\cite{big}
\begin{equation}\label{match}
P^m(S) dS = { \Gamma(N) 2^{-N+1} \over \Gamma(N/2)^2}
{(1+2\xi^2)^{N/2} \over (1+(1-S)\xi^2)^N} (1-S^2)^{N/2-1} dS \, .
\end{equation}
Here $\Gamma(z)$ is the gamma function and $\xi=\sigma_s/\sigma_\eta$
is the signal-to-noise ratio for the detector. The temperature
fluctuations in the CMBR are taken to have a gaussian temperature
distribution with variance
$\sigma_s$. If $N$ is large, the distribution is peaked at
\begin{equation}
S^m_{max} = { \xi^2 \over 1+\xi^2} + {\cal O}(N^{-1}) \, .
\end{equation}
It is considerably harder to predict the temperature distribution for
typical unmatched circles since fluctuations in the CMBR are known to
have spatial correlations. We can make a crude estimate by ignoring
these correlations and taking the temperature at each point on the sky
to be an independent gaussian random variable. The probability
distribution for unmatched circles is then\cite{big}
\begin{equation}\label{unmatch}
P^u(S) dS = {\Gamma(N)2^{-N+1} \over \Gamma(N/2)\Gamma(N/2)}
(1-S^2)^{(N-2)/2} dS\, .
\end{equation}
This distribution is centred at $S=0$ and has ${\rm
FWHM}\simeq (8\ln 2 /N)^{1/2}$. Spatial correlations will tend to move
the centre of the distribution toward larger $S$ and increase its width.
Using the distributions (\ref{match}) and (\ref{unmatch}) as a rough
guide, we see that our statistic works best if the instrument has high
resolution ($N$ large) and high signal-to-noise ($\xi$ large). 

\
\begin{figure}[h]
\vspace*{70mm}
\includegraphics{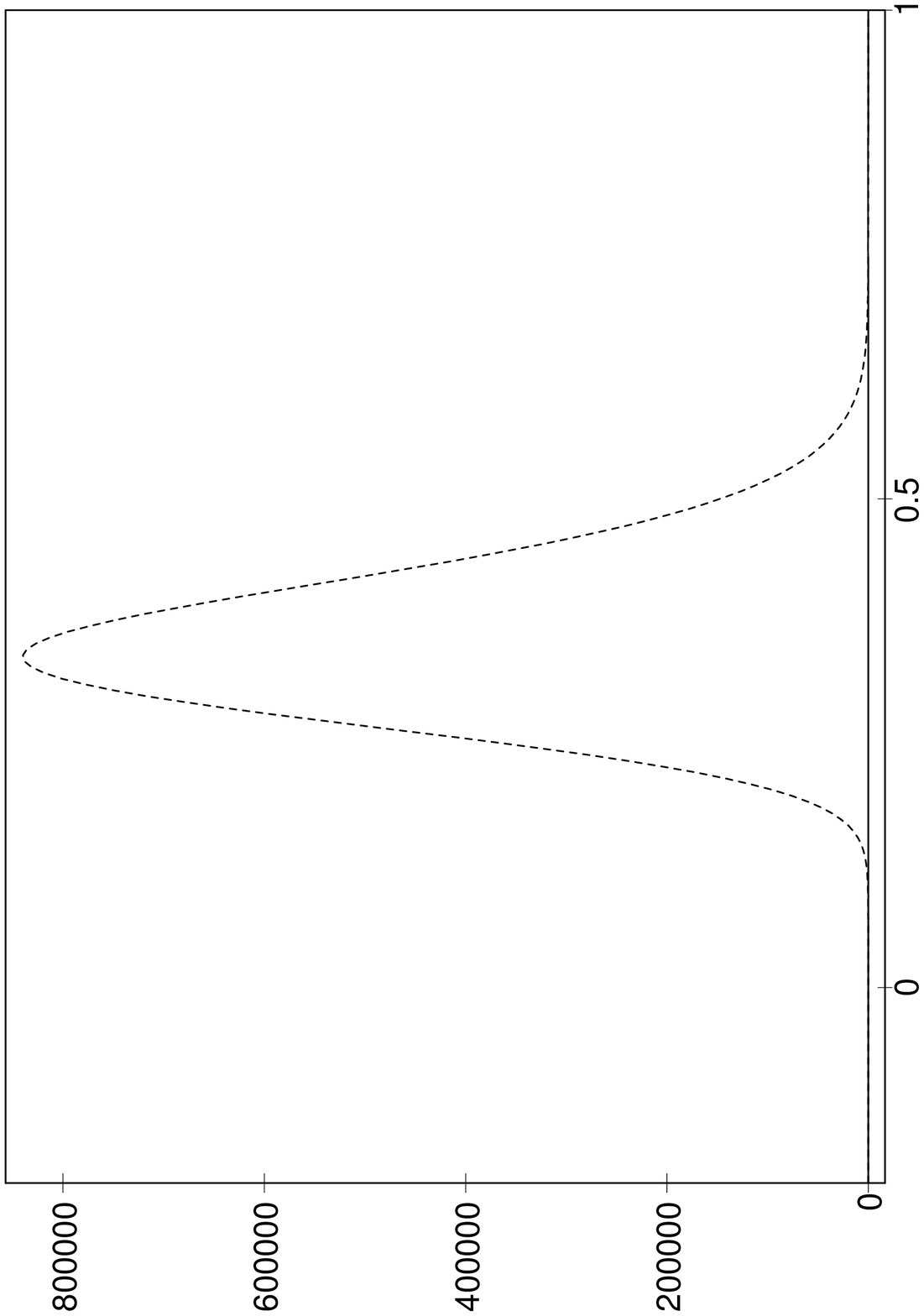}
\includegraphics{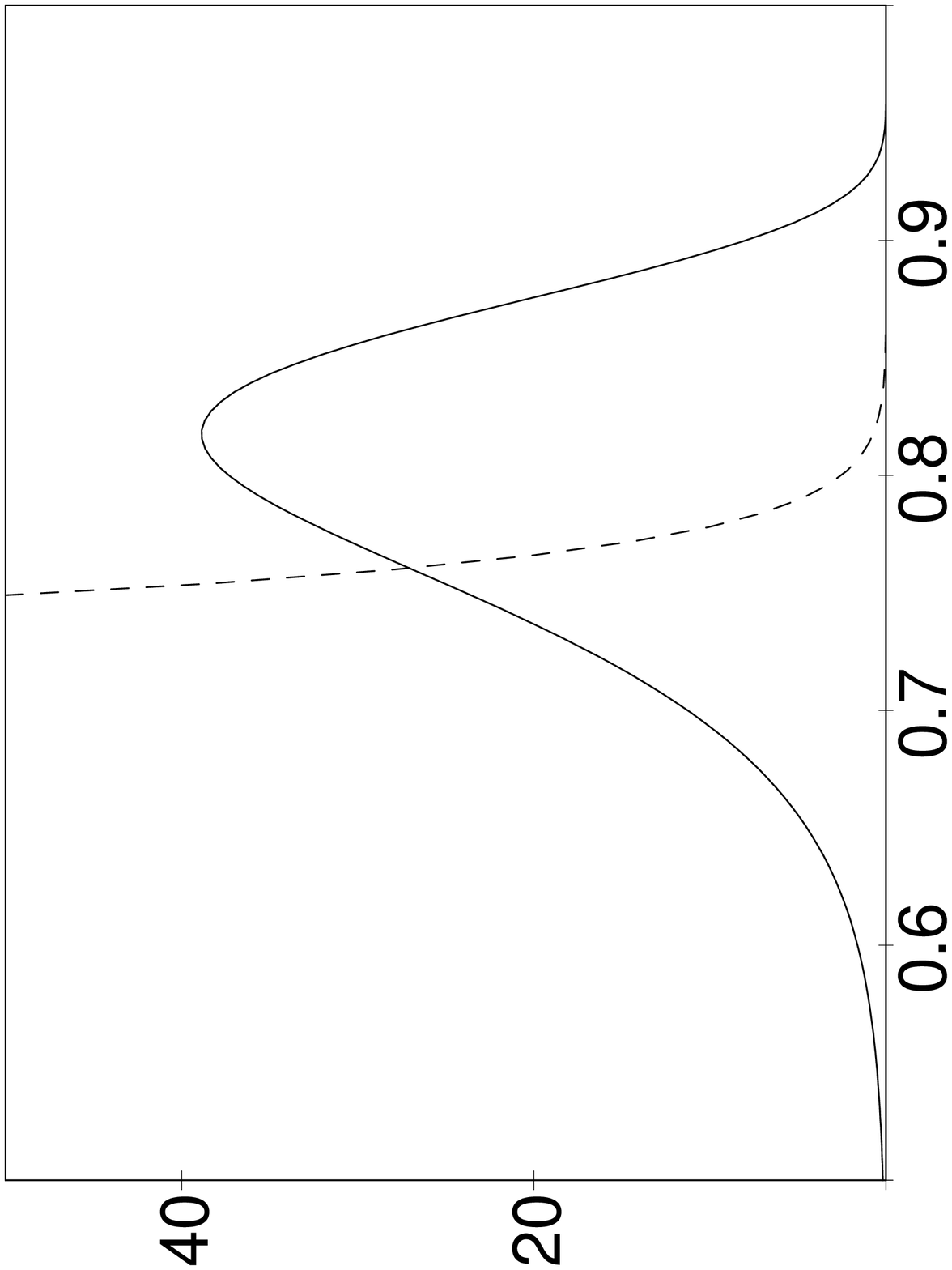}
\vspace*{9mm}
\caption{Simulated histograms for the circle statistic in a $T^3$
topology as measured by COBE-DMR. The dashed line is the histogram for
unmatched circle pairs and the solid line is histogram for the matched
circle pairs. The insert shows a detail of the region where matched
circles might be detected.}
\end{figure}

\vspace*{-6mm}
\begin{picture}(0,0)
\put(190,53){$S_{\max}$}
\end{picture}

When multiplied by the number of matched and unmatched circles pairs,
$P^m(S)$ and $P^u(S)$ can be used to produce histograms of the circle
statistic. Given an all-sky map with $M$ data points, the number of
circle pairs of a given angular radius scales as $M^2$. Moreover,
these have to be compared at $N\sim M^{1/2}$ different phasings.
There will be $\sim M^{1/2}$ sets of circles of a given angular radius
to compare, leading to a total sample that scales as $M^3$. The
storage problem this poses can be alleviated somewhat if we only store
the maximum value of $S(\phi_*)$ for each circle pair. The
distribution for unmatched circle pairs then becomes:
\begin{equation}\label{unmatch2}
P^u_{\rm max}(S_{\rm max})=N P^u(S<S_{\rm max})^{N-1} P^u(S_{\rm max}) \, .
\end{equation}
Remembering that (\ref{match}) and (\ref{unmatch2}) overestimate the
performance of the statistic, we can test to see if
the COBE-DMR 4-year data\cite{bennett} can be used to probe the
topology of the
universe. Crudely speaking, the instrument has an angular resolution
of $\Delta\theta = 10^o$ and reached a signal to noise ratio of
$\xi=2$ after collecting data for 4 years. The optimal
candidate\footnote{COBE is unable to detect models with $\Omega_0 < 1$
since fluctuations on scales larger than $\theta_{curv}$ do not
originate of the SLS and must be filtered out. In addition, while
smaller flat topologies produce more large matched circles, they are
unable to support fluctuations on the scales probed by COBE.}
topology for COBE to detect is a flat thee-torus with
topology scale roughly equal to $R_{sls}$. The simulated histograms
for this example are shown in Fig.~2. Even under the idealised
conditions we have described, it is clear that detection would at most
be marginal. The prospects are far brighter for the next generation of
satellite missions. For example, MAP will provide all-sky coverage
with a signal-to-noise of $\xi\sim 15$ at
$\Delta\theta =0.5^o$. Any matched circles in these maps will be
thrown into stark relief by our circle statistic.

Since the distribution of the circle statistic will be non-gaussian,
Monte-Carlo simulations will be necessary to
properly test the significance of a null detection or
the significance of a potential detection. We are currently simulating
the performance of the circle statistic by producing synthetic CMBR
skies in both simply and multiply connected inflationary cosmologies and
analysing them using the resolution and noise profile of the MAP
satellite. Our preliminary results are very encouraging. Our
confidence has been bolstered by the realization that the expected number of
matched circle pairs for a generic small topology is very large 
(see section 4), and by Weeks' observation\cite{jeff} that just a few
matched circle pairs can be used to find the generators of a
topology's fundamental group. This allows the entire topology to be
reconstructed and the size and position of all the other circle pairs
to be predicted. If these predictions agree with the data, then there
will be absolutely no doubt about the result.

\section{Circles in a $T^3$ universe}

It is particularly easy to illustrate some of our ideas in
a flat universe with three-torus topology. Using
methods similar to those described in Ref.\cite{SSS}, we have simulated
the CMBR in a cubic $T^3$ with topology scale $L=R_{sls}$. Our
simulations were run at a resolution of $\Delta\theta=1.4^o$ using a
flat Harrison-Zeldovich power spectrum and variable signal-to-noise.
In Fig.~3 we display front and back views of the sphere of last
scatter. Also marked are one pair of matched circles with angular
radii of $\alpha=59.8^o$. Matched pairs were found by
evaluating the circle statistic $S$ for all $2\times 10^{10}$ distinct
circle pairs with angular radii in the range $10^o\leq \alpha \leq 90^o$.
The number of distinct circle comparisons is finite as the search
increment is fixed by the angular resolution of $\Delta\theta=1.4^o$.

In Fig.~4 we display the variation in CMBR temperature around each
circle. Since the match is so good, we also show
the temperature difference $T_1(\phi)-T_2(\phi)$. There is no phase
offset for this matched pair. The graphs in Fig.~4 are drawn without
noise. The noise is added when we simulate the detector response.
In theory the match should be exact. The discrepancy is caused by
what we call ``pixel noise''. Since the

\newpage

\
\begin{figure}[h]
\vspace*{75mm}
\includegraphics{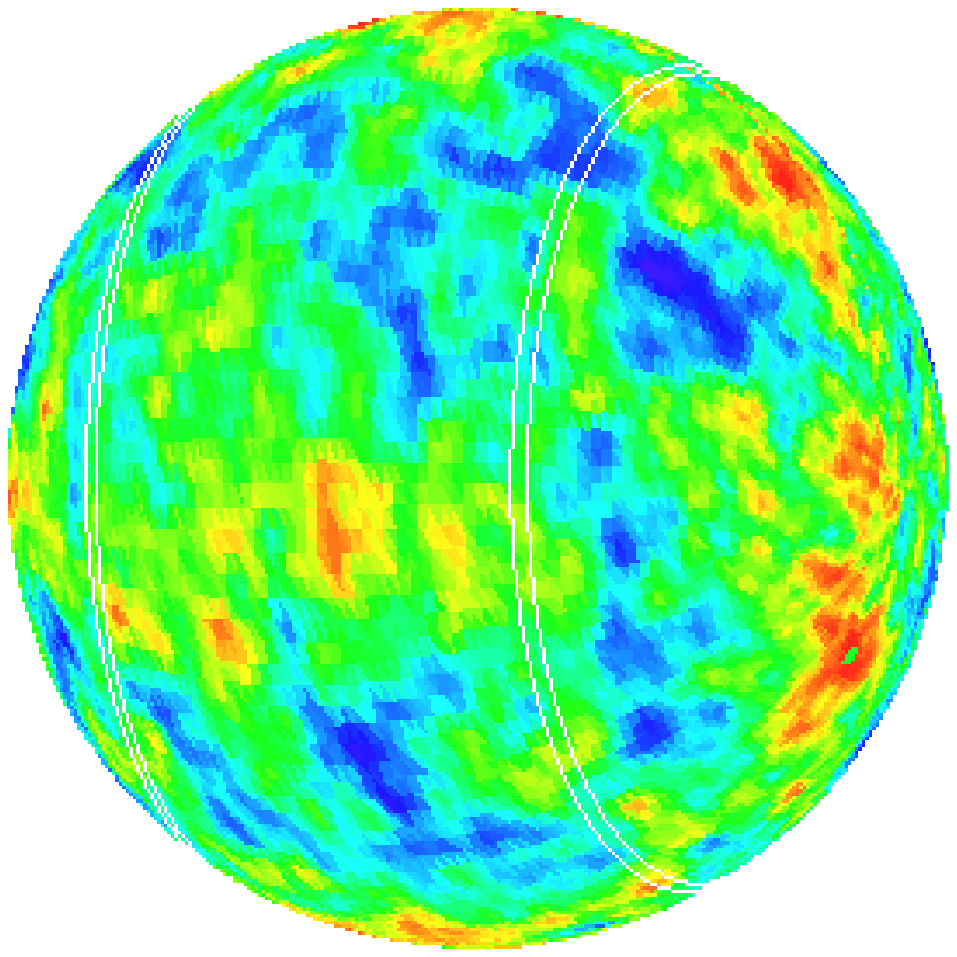}
\includegraphics{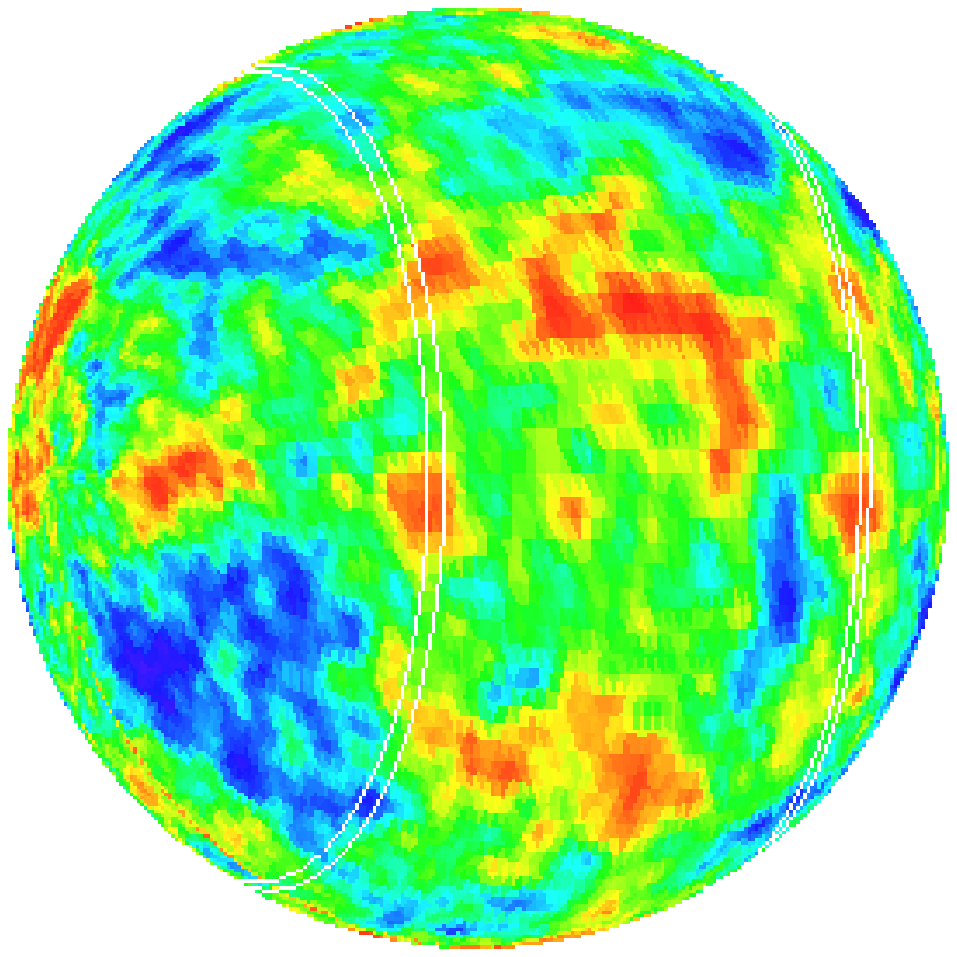}
\includegraphics{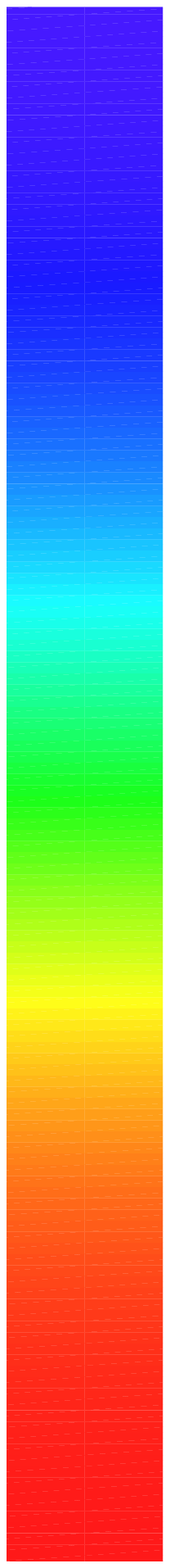}
\caption{Views from opposite sides of the SLS globe. Two matched
circles are marked.}
\end{figure}

\begin{picture}(0,0)
\put(108,60){-3}
\put(269,60){3}
\put(185,59){{ ${T}$}}
\end{picture}

\
\begin{figure}[h]
\vspace*{70mm}
\includegraphics{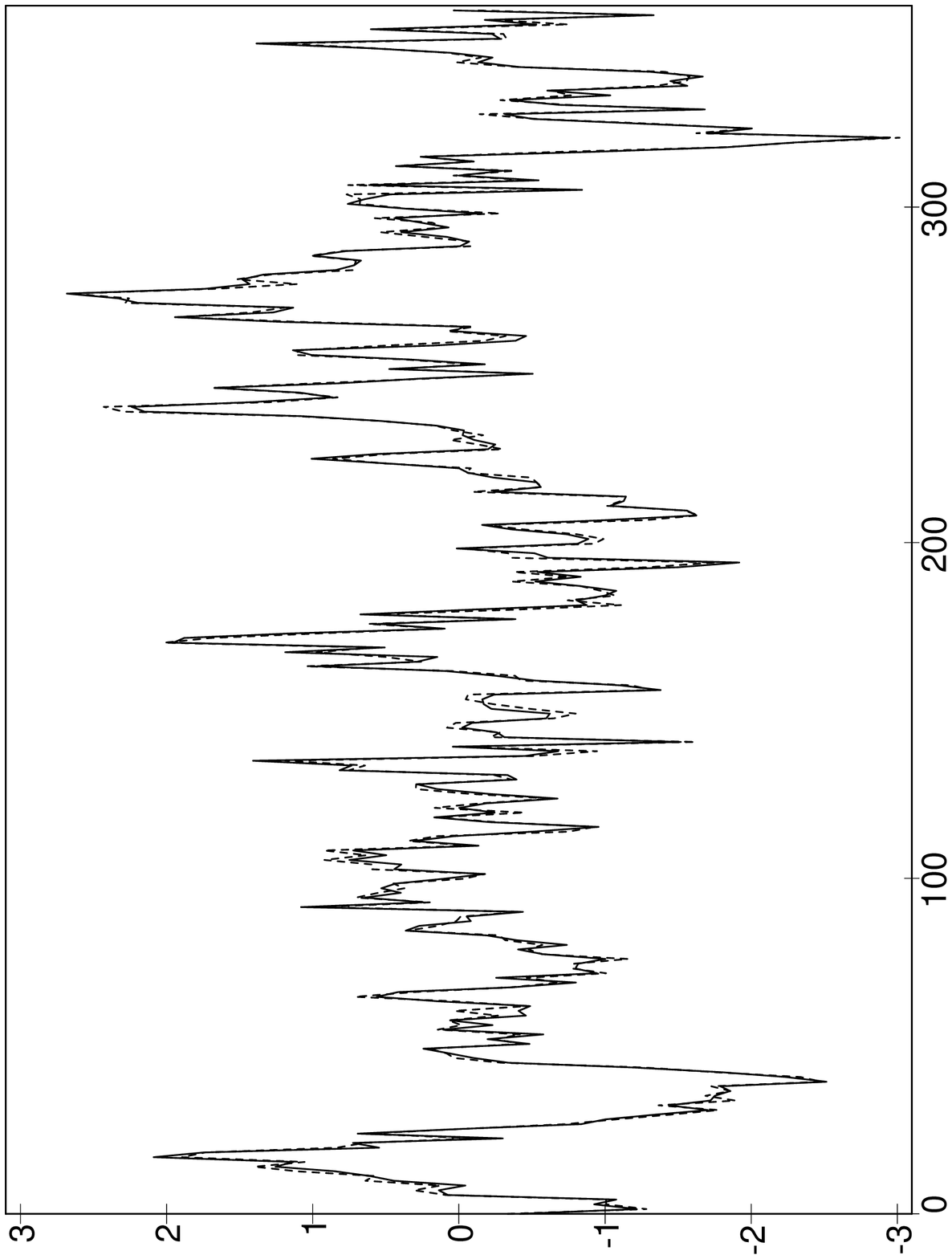}
\includegraphics{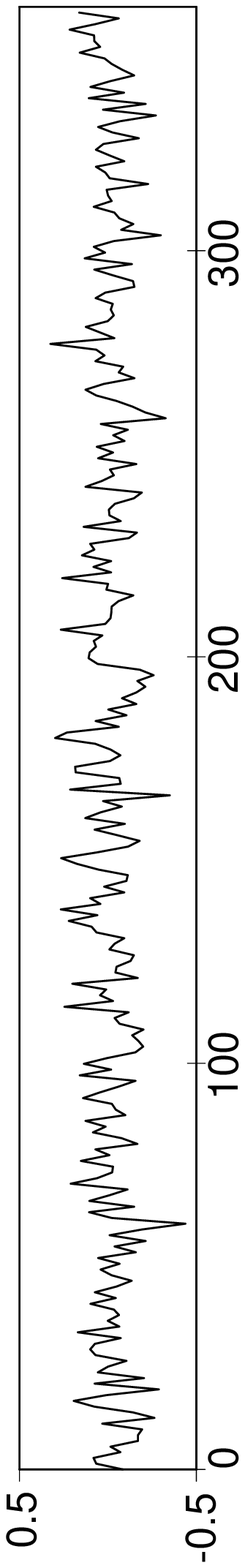}
\vspace*{15mm}
\caption{(a) The CMBR temperatures around the two circles. The solid
line is $T_1(\phi)$ and the dashed line is $T_2(\phi)$. (b) The
temperature difference $T_1(\phi)-T_2(\phi)$.}
\end{figure}

\vspace*{-5mm}
\begin{picture}(0,0)
\put(45,270){(a)}
\put(45,75){(b)}
\put(40,190){$T(\phi)$}
\put(180,40){{ $\phi$}}
\end{picture}

\noindent angular resolution is
finite, the circles we compare do not have exactly the right centrings or
angular radii. Moreover, the temperature at each point around a circle
is found by linear interpolation from the sky pixel
temperatures. These effects combine to produce pixel noise. The pixel
noise is reduced if we go to higher angular resolution. In addition,
experimental data sets are usually oversampled, {\it i.e.} the data is
recorded at an angular resolution substantially higher than the
detector resolution. For COBE the data was collected at a resolution
of $2.4^o$ while the true instrument resolution was $10^o$. Because of
this oversampling, there is no information in the temperature gradient
between adjacent pixels. Consequently, there will be no pixel noise
coming from our temperature interpolation when we apply our search to
real data sets.

The pair of matched circles have exactly the angular size and relative
position we expect from the intersection of the sphere of last scatter
with its nearest clone. In a flat universe the angular radius of a
matched circle pair follows from simple trigonometry:
\begin{equation}\label{flatangle}
\alpha = {\rm arccos}\left( { X \over 2 R_{sls} } \right) \, .
\end{equation}
Here $X$ is the distance between us and the clone responsible for the
matched circle. For a cubic $T^3$ with sidelength $L=R_{sls}$ there
will be matched circle pairs with $\alpha=\{ 60^o, 45^o, 30^0, 0^o\}$
and multiplicities $\{3,6,4,3\}$. 

\
\begin{figure}[h]
\vspace{55mm}

\includegraphics{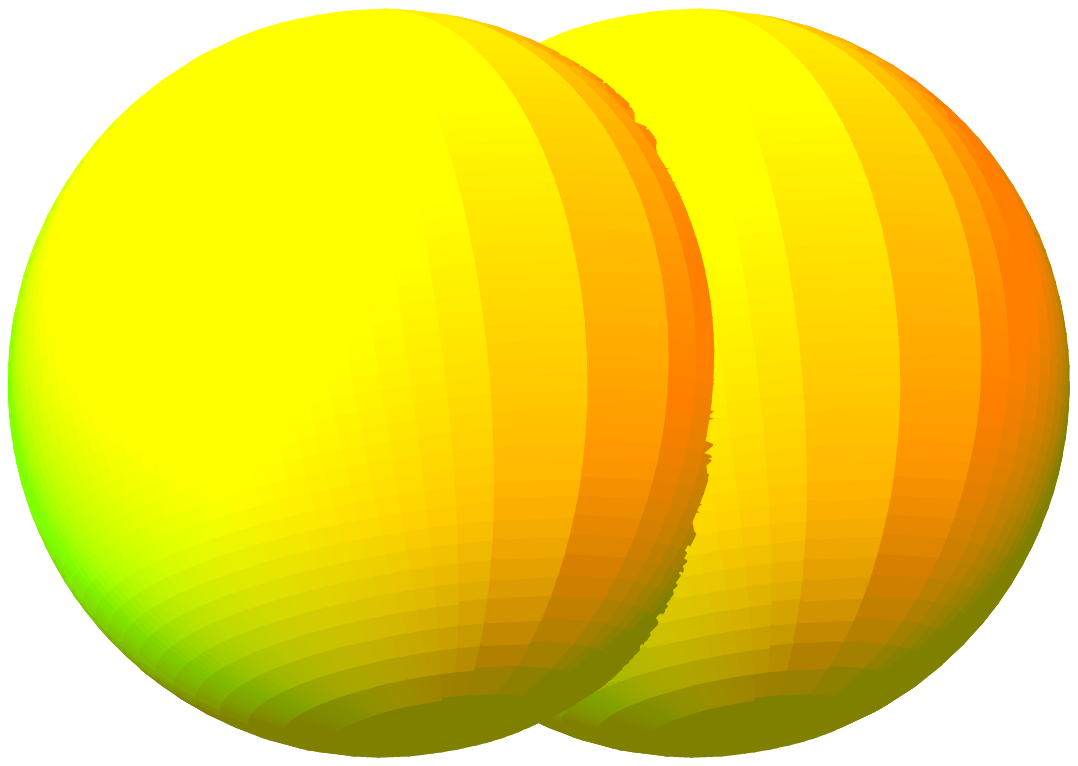}
\includegraphics{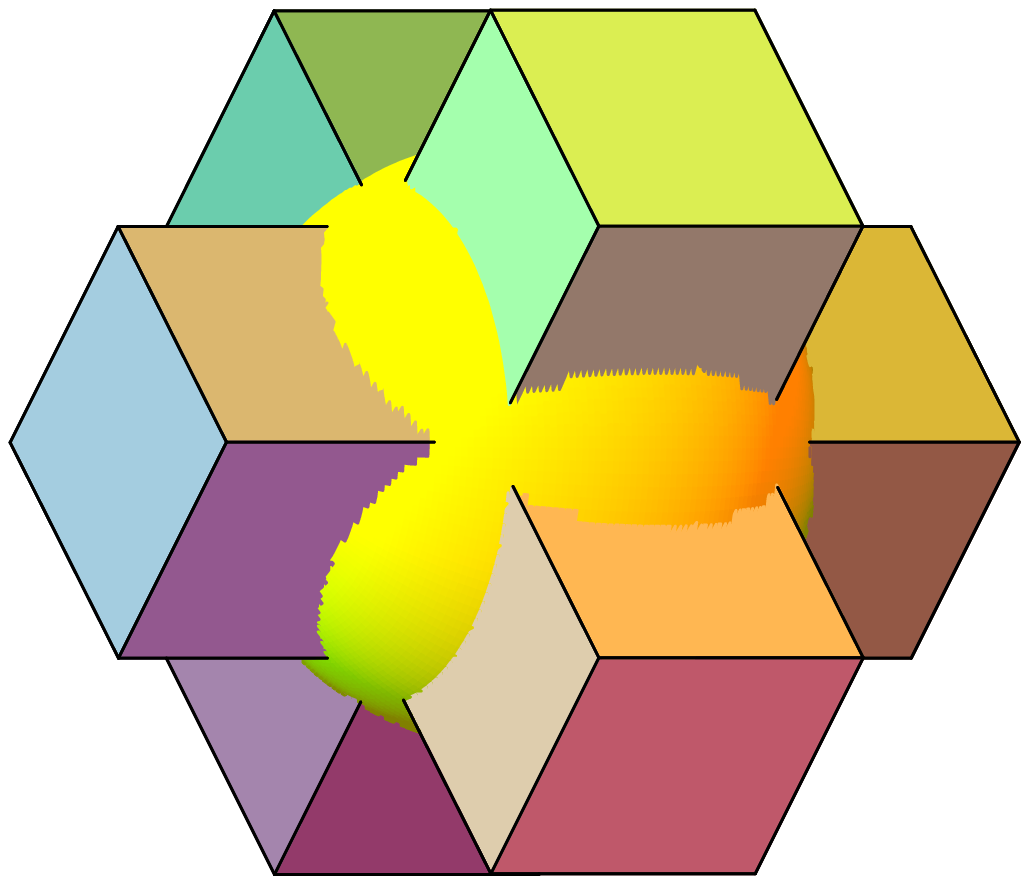}

\vspace{5mm}
\caption{(a) The intersection that leads
to the matched circle pair (b) The sphere of last scatter centred on
the origin and the six nearest neighbours to the central fundamental cell.}
\end{figure}

\vspace*{-2mm}
\begin{picture}(0,0)
\put(40,190){(a)}
\put(200,190){(b)}
\end{picture}

The matched circles seen in Fig.~3 are due
to the intersection shown in Fig.~5a. A partial tiling of the covering
space by our six nearest neighbour fundamental cells is shown in
Fig.~5b. The matched circles arising from these clones lie on faces of
the fundamental cell. However, this is not true for all matched
circle pairs. For example, the circles coming from the clones situated in
the next-to-nearest neighbour cells (with $\alpha=45^o$) do not lie on
faces of the fundamental cube.

We can evaluate the circle statistic $S(\phi_*)$ for the matched pair
under consideration. This is shown in Fig.~6 as a function of pixel
offset $i= \phi_* / 2\Delta \theta$. There is a clear peak
at $\phi_*=0$ where the the statistic
reaches the value $S(0)=0.975$. This should be compared to the FWHM
of $S=0.21$ predicted by (\ref{unmatch}) for the distribution
of unmatched circle pairs with the same radius. We are in the
process of producing
histograms for the complete search to confirm there is a clear
detection. The results produced here were obtained by setting a
threshold of $S_{cut}=0.95$ and only keeping circle pairs with
$S>S_{cut}$. It was encouraging to find that no topologically
unmatched pairs made the cut.

\vspace*{65mm}
\begin{figure}
\includegraphics{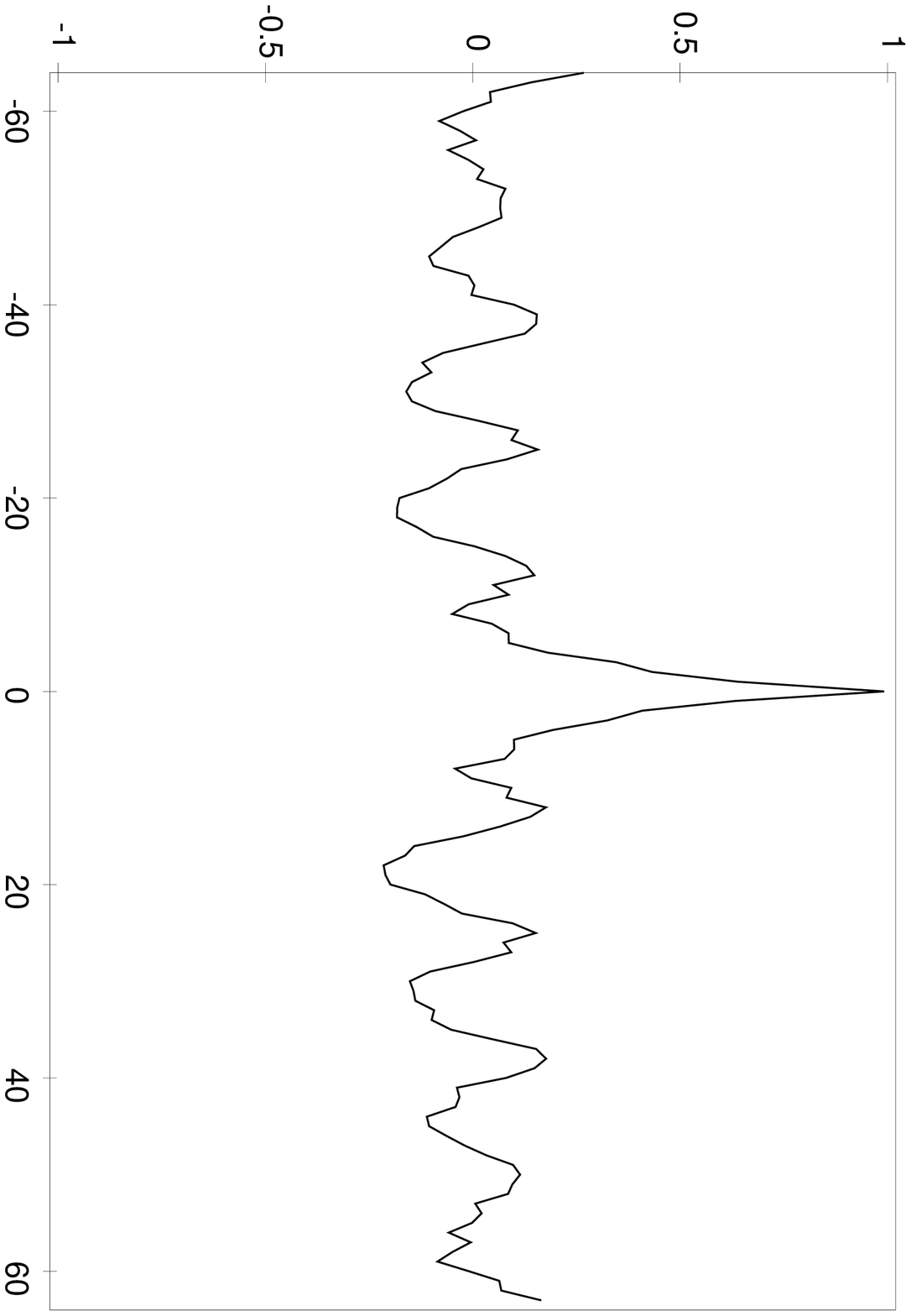}
\vspace*{5mm}
\caption{The circle comparison statistic, $S$, versus the pixel offset
of the matched circle pair}
\end{figure}

\vspace*{-4mm}
\begin{picture}(0,0)
\put(50,142){$S$}
\put(195,38){{ $i$}}
\end{picture}

\
\begin{figure}[h]
\vspace*{60mm}
\includegraphics{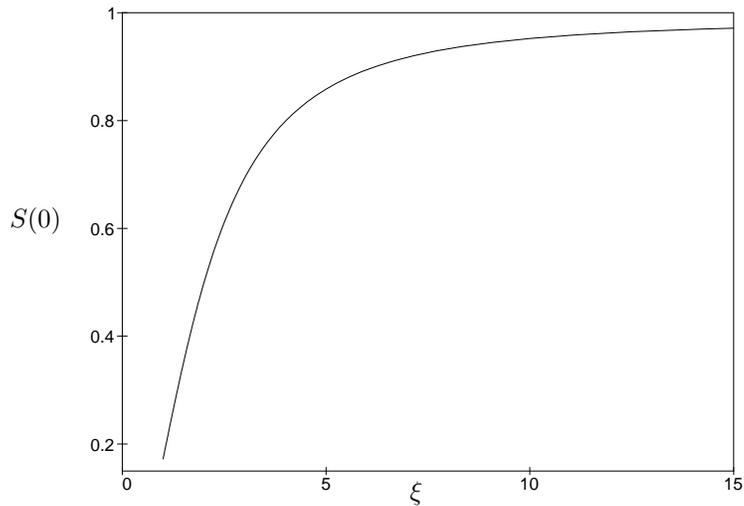}
\vspace*{5mm}
\caption{The degradation in the circle match caused by detector noise.}
\end{figure}

\vspace*{-2mm}
\begin{picture}(0,0)
\put(35,143){$S(0)$}
\put(183,40){{ $\xi$}}
\end{picture}

We are also studying how other
complications such as detector noise and the galaxy cut affect our
ability to detect topology. In Fig.~7 we show how detector noise
degrades the match.
The graph suggest that a signal-to-noise of at least $\xi=5$ is
required to get a good match. The next generation of CMBR
satellite missions will far exceed this requirement.

\section{Circles in a compact hyperbolic universe}

The real power of the circle test is revealed when we apply it to models
with compact hyperbolic spatial sections\cite{cornish}. In these
models it is very difficult to predict what the CMBR temperature
fluctuations should look like as the inflationary perturbations
are described by non-analytic functions\cite{cobepaper}. This makes
it difficult to place constraints on compact hyperbolic models using
methods based on comparisons between predicted and observed CMBR
maps\cite{bond,janna}. In contrast, the circle test requires no
knowledge about what caused
the temperature fluctuations on the sphere of last scatter. They could
have been painted on by elves for all we care. The only difficulty we
encounter is in trying to perform Monte-Carlo simulations of the
circle search as these require synthetic sky maps.

\
\begin{figure}[h]
\vspace{60mm}

\includegraphics{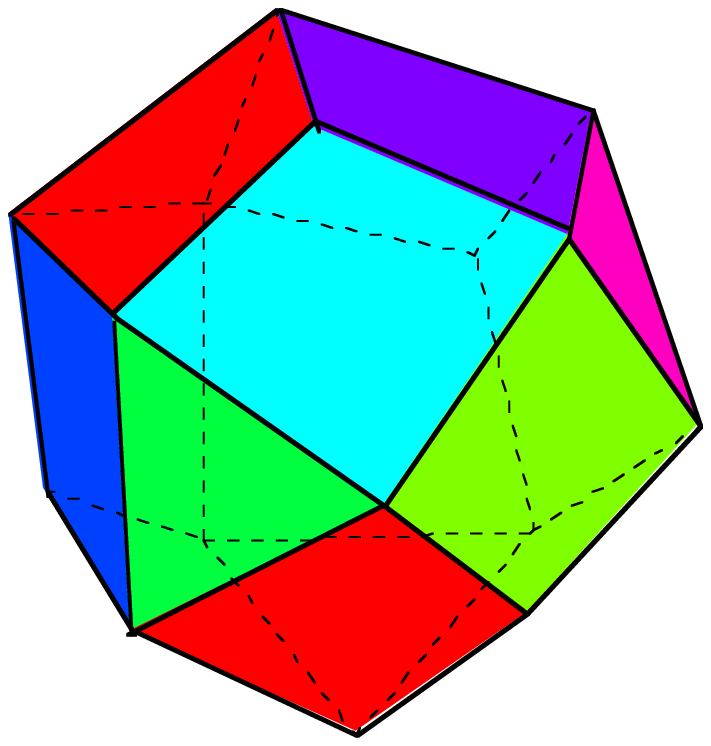}
\includegraphics{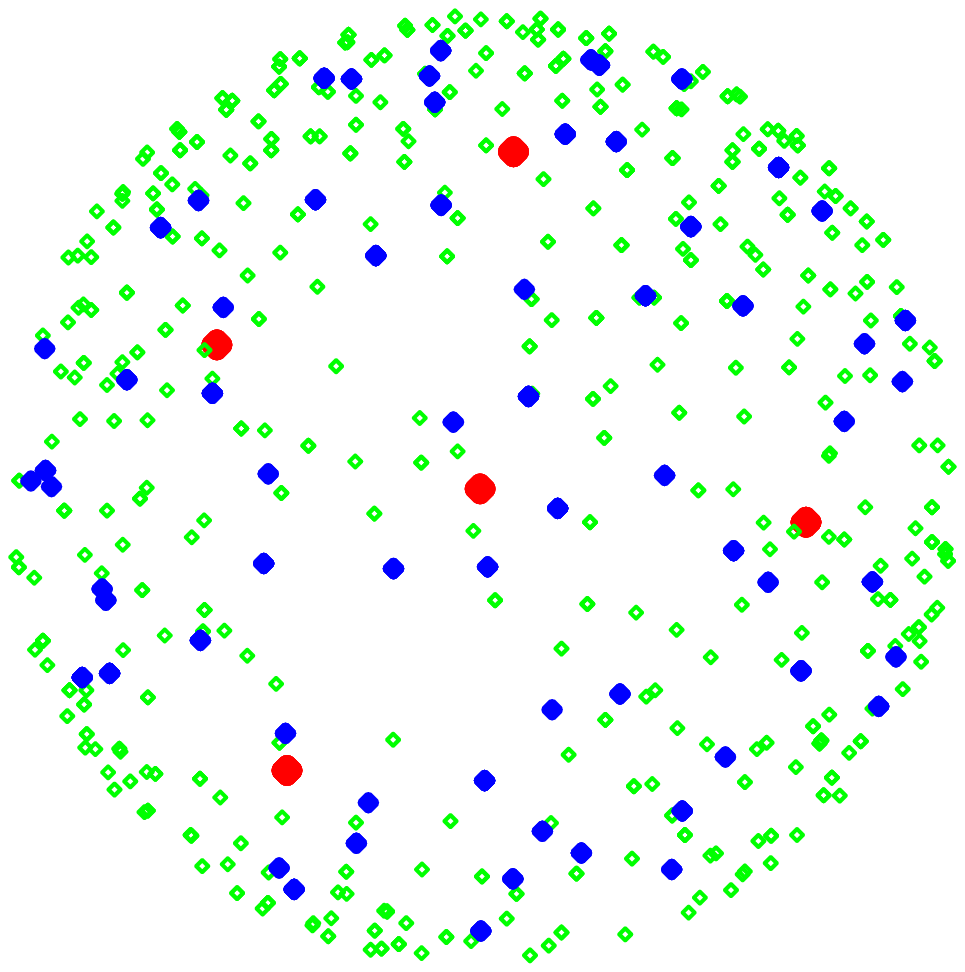}

\vspace{2mm}
\caption{The distribution of our clones in the Thurston universe out
to a radius of $3 R_{curv}$. The large points are within one
curvature radius, the medium sized points are within two
curvature radii and the small points are within three curvature
radii. The fundamental cell centred on the origin is also shown.}
\end{figure}

What we can do fairly easily is make some predictions about the
number, size and distribution of circles in a generic small hyperbolic
universes. The number, ${\cal N}$, of
topologically matched circle pairs is given
by the number of our clones within a proper distance of
$D_{sls}$. A good estimate for this number is given by the ratio of
the volume of space enclosed by a ball of radius $D_{sls}$ and the
volume of the topology's fundamental cell:
\begin{equation}
{\cal N}={ \pi (\sinh(2 D_{sls}/R_{curv}) - 2 D_{sls}/R_{curv}) \over 
{\rm Vol}(\Sigma) } \, .
\end{equation}
Once the radius of the ball exceeds the curvature radius, the volume
of the ball, and hence the number of matched circles, grows
exponentially with increasing radius. This makes the number of matched
pairs a very sensitive function of the energy density. In Fig.~8 we
show the distribution of clones in a model of the universe where the
spatial sections are those of Thurston's manifold\cite{bills}. The picture is
drawn using Klein's projective model of hyperbolic space, showing all
clones out to a radius of $3 R_{curv}$.

There are a two things to notice about this picture. The first is that
most of our clones are at least $2R_{curv}$ away from the origin. The
second is that the distribution is fairly isotropic (though this is
not so evident in our 2-dimensional rendering of the 3-dimensional
distribution). Using the fact that Thurston's manifold has volume
$0.981 R_{curv}^3$, and recalling that
$R_{sls}= R_{curv}{\rm arccosh}[(2-\Omega_0)/\Omega_0]$, it is possible to estimate
how the number of matched circles varies with the density parameter.
This is shown in Fig.~9. Clearly, our chances of finding matched circles
increases rapidly as the density of the universe decreases. In the
observationally favoured range of $\Omega_0= 0.3 \rightarrow 0.4$, we
expect to see tens of thousands of matched circles if we live in
Thurston's universe. Even if the fundamental cell has a volume as
large as $1000 R_{curv}^3$, we might still discover the topology
of the universe.

\begin{figure}[h]
\vspace*{55mm}
\includegraphics{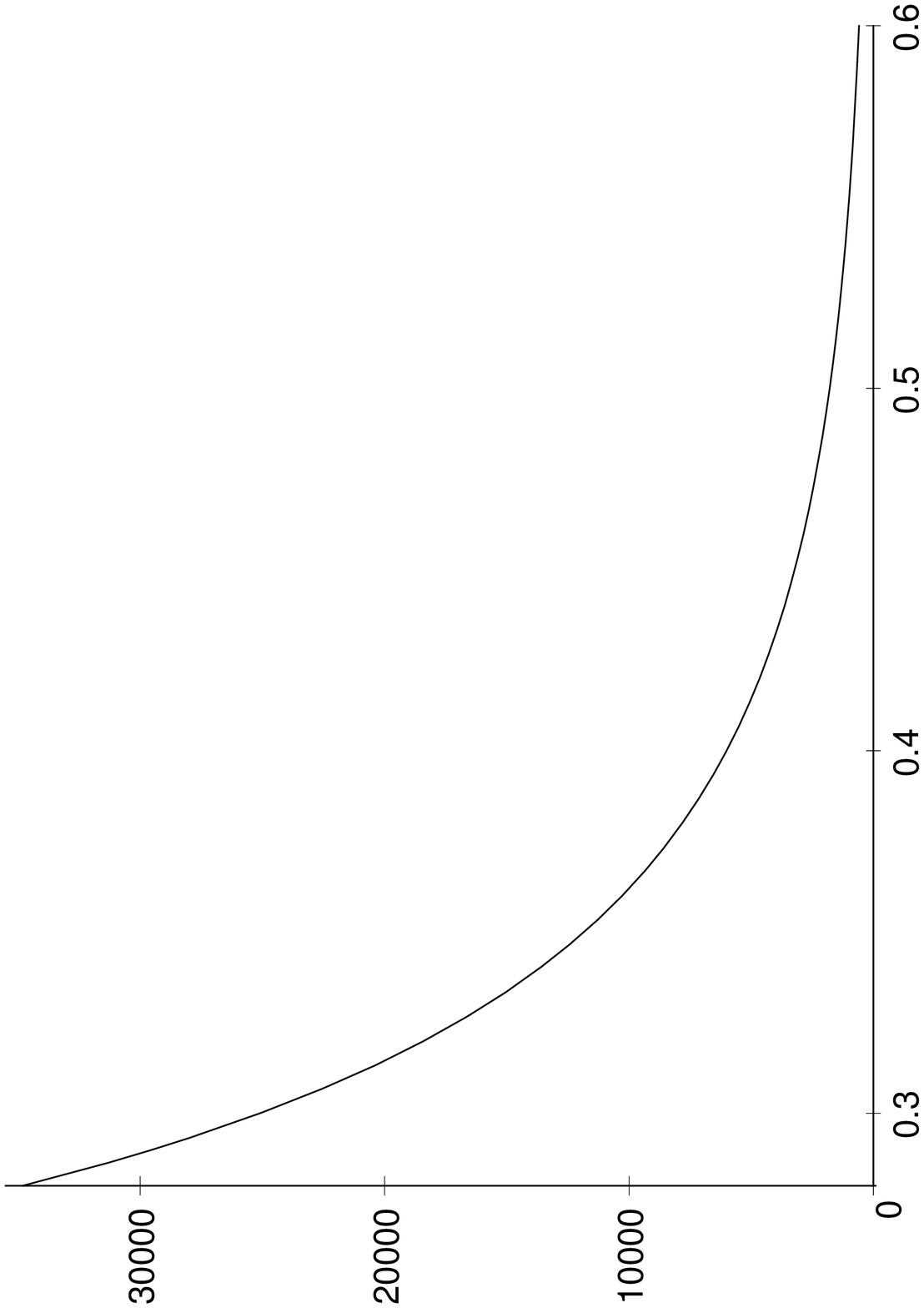}
\vspace*{5mm}
\caption{The number of matched circles in Thurston's universe as a
function of the density parameter.}
\end{figure}

\vspace*{-5mm}
\begin{picture}(0,0)
\put(55,120){${\cal N}$}
\put(190,32){{ $\Omega_0$}}
\end{picture}

Using a little hyperbolic trigonometry we can predict the angular
radius of the circles as a function of the proper distance $X$ between
us and the clone responsible for the match:
\begin{equation}
\alpha = {\rm arccos}\left( { \cosh(X/R_{curv})-1 \over
\sinh(X/R_{curv}) \tanh( R_{sls}/R_{curv}) } \right).
\end{equation}
In the limit $R_{curv}\rightarrow \infty$ we recover the flat space
result (\ref{flatangle}). Since most of our clones will be situated
near the boundary of the region $X < D_{sls}$, most matched circles
will have small angular radii. In Fig.~10 we plot histograms showing
the distribution of matched
circle radii for Thurston's universe with $\Omega_0=0.3$ and
$\Omega_0=0.5$. The angular bins were taken to be one degree wide.

The circle statistic (\ref{cstat}) works best when the number of
pixels, $N$, around each circle is large. Since
$N = 360 \sin\alpha /\Delta\theta$, the test works best for large circles. 
Our preliminary results indicate that we need $N \gtrsim 50$ to make a clear
detection from a sky map with $\xi \sim 10$. Taking
$\Delta\theta=0.5^o$, this tells us that only for $\alpha>4^o$ can we
confidently identify individual circle pairs.  The histograms in
Fig. 10 suggest that this restriction would not seriously harm
our ability to probe the topology of the universe.
Moreover, because we expect many pairs of smaller circle pairs,
and because, as pointed out by Weeks, any subset of two to three of
them can be used to predict the location of all the others, we can use
the self-consistency conditions to eliminate false positive identifications.
We therefore expect to be able to probe somewhat, and perhaps
substantially, below $\alpha=4^o$.

\begin{figure}[h]
\vspace*{55mm}
\includegraphics{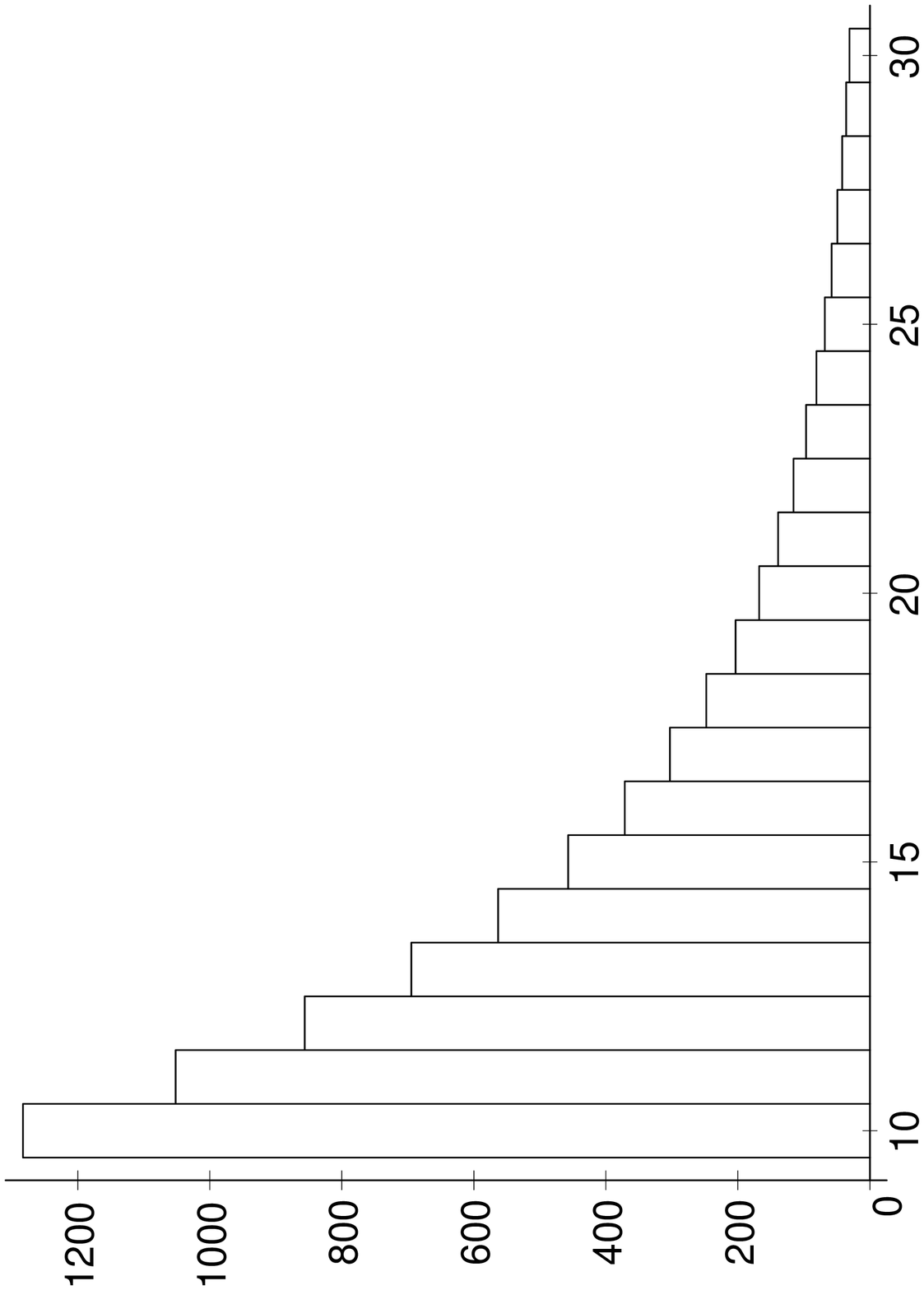}
\includegraphics{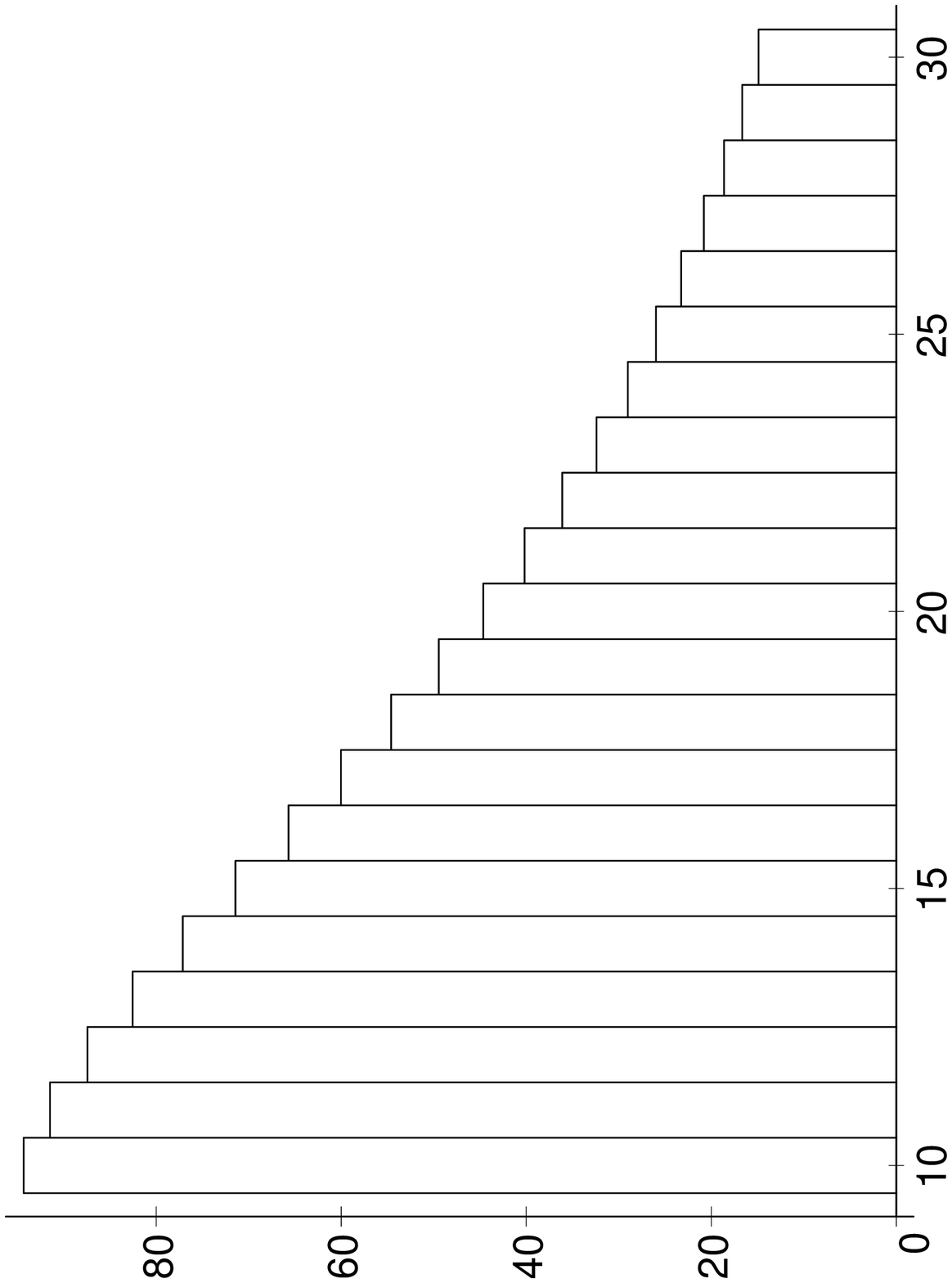}
\vspace*{5mm}
\caption{The distribution of matched circles as a function of angular size.}
\end{figure}

\vspace*{-5mm}
\begin{picture}(0,0)
\put(100,150){$\Omega_{0}=0.3$}
\put(300,150){$\Omega_{0}=0.5$}
\put(110,27){{ $\alpha$}}
\put(290,27){$\alpha$}
\end{picture}

Compact hyperbolic models provide our best chance to discover
the topology of the universe. Compared to a small flat universe of
similar size, the exponentially larger volume of hyperbolic space
leads to many more matched circles. While most of these circles have
small angular radii, there should be enough large circles to make
detection possible.

\section{Prospects}

The COBE 4-year map will not be our ultimate map of
the microwave sky.  NASA is currently making preparations to launch the MAP
satellite in the year 2000\cite{map}. MAP represents a ten-fold improvement in
signal-to-noise and more than a thirty-fold improvement in
resolution over the COBE-DMR map. ESA is also planing the
PLANCK mission, to be launched in 2006\cite{planck}. PLANCK will have
even higher resolution than MAP.

The topological signatures that we plan to search for should be
easily detectable (if they are present) in the 
higher resolution, lower noise maps that will soon be available.
If we find generic signals of topology, we will
be able to identify the particular topology in which we live\cite{jeff}, 
where we are within the topology and which way is up.
Using our synthetic sky maps, we will develop the technology
to identify individual topologies in current and future data
samples. 

In summary, the possibility of non-trivial topology greatly
widens and enriches the zoo of possible cosmologies.  
We have suggested that for generic
small universes the ideal signal to look for is topologically
identified circle pairs in the microwave background. We have devised a
statistic to test for matched circle pairs and a strategy for
performing the search. We are currently road-testing our search
programs on simulated CMBR maps and hope to report our findings in
the near future\cite{big}.

If we do detect the signature of finite topology, its implications
would be profound and have great popular interest:
we would learn that it's small universe after all.

\end{document}